\documentclass[fleqn,10pt]{wlscirep}

\usepackage{graphicx}  
\usepackage{dcolumn}   
\usepackage{bm}        
\usepackage{amssymb}   
\usepackage{soul}
\usepackage{mathtools}
\usepackage{doi}

\usepackage{tikz}
\usetikzlibrary{arrows}
\usetikzlibrary{positioning,decorations.pathreplacing}
\usetikzlibrary{shapes.multipart}

\usepackage{ulem}
\usepackage{caption}
\usepackage{subcaption}
\usepackage{graphicx}
\usepackage{stackengine}

\newcommand*{\bfrac}[2]{\genfrac{\lbrace}{\rbrace}{0pt}{}{#1}{#2}}
\newcommand\myabove[2]{\genfrac{}{}{0pt}{}{#1}{#2}}

\usepackage{amsmath}
\newcommand{\e}{\epsilon}

\newcommand{\de}{\Delta\epsilon}
\newcommand{\te}{\tilde{\e}}
\newcommand{\tG}{\tilde{\Gamma}}
\newcommand{\dU}{\Delta U}
\newcommand{\w}{\omega}

\begin{document}



\title{Quantum phase transition in a realistic double-quantum-dot system}
\date{\today}
\date{\today}

\author[1,*]{Yaakov Kleeorin}
\author[1,2]{Yigal Meir}

\affil[1]{Department of Physics, Ben-Gurion University of the Negev, Beer Sheva 84105, Israel}
\affil[2]{The Ilse Katz Institute for Nanoscale Science and Technology, Ben-Gurion University of the	Negev, Beer Sheva 84105, Israel}

\affil[*]{kleeorin@post.bgu.ac.il}


\begin{abstract}
Observing quantum phase transitions in mesoscopic systems is a daunting task, thwarted by the difficulty of experimentally varying the magnetic interactions, the typical driving force behind these phase transitions. Here we demonstrate that in realistic coupled double-dot systems, the level energy difference between the two dots, which can be easily tuned experimentally, can drive the system through a phase transition, when its value crosses the difference between the intra- and inter-dot Coulomb repulsion. Using the numerical renormalization group and the semi-analytic slave-boson mean-field theory, we study the nature of this phase transition, and demonstrate, by mapping the Hamiltonian into an even-odd basis, that indeed the competition between the dot level energy difference and the difference in repulsion energies governs the sign and magnitude of the effective magnetic interaction. The observational consequences of this transition are discussed.

\end{abstract}

\maketitle
\section*{Introduction}
Quantum phase transitions (QPTs), where a system changes its zero-temperature phase when a physical parameter is continuously varied, is one of the focal research areas in physics in general and condensed matter physics in particular \cite{Sondhi1997,Sachdev2011}. While usually QPTs are associated with bulk systems, they may be observed even in mesoscopic systems. One simple example is a single Kondo impurity - a magnetic moment embedded in a Fermi sea \cite{Hewson}. As was pointed out by Anderson \cite{Anderson1970c}, the system changes its character from a Kondo state, where the electrons in the Fermi sea screen the impurity, to a trivial state, where the magnetic interaction $J$ between the impurity and the electrons renormalizes to zero, as the $z$-component of $J$ changes from being anti-ferromangetic (AFM) to ferromagnetic (FM). Similarly, in a double-impurity system, one expects, as the AFM interaction between the two impurities is decreased, a transition from a state where  this interaction dominates,  yielding a zero-spin entity that is decoupled from the electrons, to a state where each impurity is Kondo screened by the Fermi sea \cite{Jones1988}. In addition, a phase transition has also been predicted as the double dot system is driven from a triplet to a singlet ground state as $J$ changes sign, similarly to the single-dot case \cite{Vojta2002,Hofstetter2002a}. While Kondo physics has been predicted \cite{Glazman1988a,Ng1988,Meir1993b} and observed \cite{Goldhaber-Gordon1998b,Cronenwett1998b} in quantum dots (QDs), observing such phase transitions in these highly controllable systems proved a daunting task, mainly due to the fact that  it is very hard experimentally to continuously modify the magnetic interactions in the system. However, a transition between singlet and triplet ground states of a single dot can be induced experimentally by a magnetic field \cite{Sasaki2000,VanderWiel2002} or by changing the effective potential \cite{Kogan2003}, leading to a crossover from a non-Kondo to a Kondo regime.

In this paper we demonstrate that a realistic device, consisting of two QDs with different energies and with different inter-dot and intra-dot Coulomb interactions, coupled in parallel to two leads, displays a QPT by simply gate tuning the on-site energy difference between the two QDs at the chemical potential corresponding to the particle-hole (PH) symmetric point, where the two-dot system is doubly occupied. (Similar setups have already been studied, but with no energy difference between the dots\cite{Wang2011a} or in the absence of inter-impurity repulsion\cite{Xiong2017}, both of which play an important role in our formulation.)
After demonstrating the transition numerically, employing the numerical-renormalization-group (NRG) method, and semi-analytically, using slave-boson mean field theory (SBMFT), we show, by transforming the system Hamiltonian to an even-odd basis, that the difference between the dot energies, relative to  the difference between the inter- and intra-dot repulsions, plays the role of a magnetic interaction, which changes its sign, from FM to AFM, at the point where the QPT takes place.
We also discuss the nature of these two phases.

\section*{Model}
The Hamiltonian that describes the two-QD system, depicted in the inset to Fig.\ref{fig:DOS2d}, is given by
\begin{equation}
H_{2QD}=\sum_{m \sigma} \e_{m} \hat{n}_{m \sigma} + \sum_{m} U_m \hat{n}_{m\uparrow}\hat{n}_{m\downarrow} + U_{12} \hat{n}_1 \hat{n}_2
\label{eq:H2d}
\end{equation}
where $m=1,2$ denotes the QD index, $\hat{n}_{m \sigma}=d^\dagger_{m \sigma} d_{m \sigma}$, $\hat{n}_m=\sum_{\sigma} \hat{n}_{m \sigma}$ ($d^\dagger_{m \sigma}$ creates an electron in QD $m$ with spin $\sigma$), and spin-degeneracy has been assumed (i.e. no magnetic field). In order to reduce the number of parameters, we assume the same intra-dot interaction on both dots, $U_1=U_2=U$, though this assumption is not necessary, and our results also hold when these energies are different. Without loss of generalization, we assume $\e_1\le\e_2$.  We also assume that $U_{12}<U$, as one would expect experimentally. Under these assumptions, the six two-electron states have 3 distinct energies: the 4 states $|\sigma,\sigma'>$, where each dot is occupied by a single electron (of spins $\sigma$ and $\sigma'$, respectively) are degenerate with energy $2\e_1+\de+U_{12}$, where
$\de\equiv\e_2-\e_1$, while the state $|\uparrow\downarrow,0>$ where the two electrons occupy the dot with the lower energy state, has energy $2\e_1+U$. The state $|0,\uparrow\downarrow>$ is always higher in energy since $\de\ge0$ and $\dU\equiv U-U_{12}>0$. Thus, as $\de$ increases from zero, the degeneracy of the ground state changes from being 4 to 5, for $\de=\dU$, and then to a non-degenerate ground state for larger $\de$. It is the transition around this special point that we concentrate upon in this paper.
The full Hamiltonian of the double-QD system, connected in parallel to a single channel in the leads, is then given by
\begin{equation}
H = H_{2QD}\ + \sum_{\sigma k\in L,R} \epsilon_{k}c^\dagger_{k\sigma} c_{k\sigma}+ \sum_{\myabove{m \sigma}{k\in L,R}} \left(V_{mk}d^\dagger_{m \sigma} c_{\sigma k } + h.c\right) ,
\label{eq:H}
\end{equation}
where $c^\dagger_{k\sigma}$ creates an electron with spin $\sigma$ in the left (L) or right (R) lead in  momentum state $k$. For simplicity,  the tunneling amplitude is chosen to be momentum, site and spin independent,   $V_{mk}=V$.

\section*{Results}
\subsection*{Numerical Renormalization Group}
\begin{figure}[h]
	\def\big{\includegraphics[width=1\textwidth,height=0.32\textwidth]{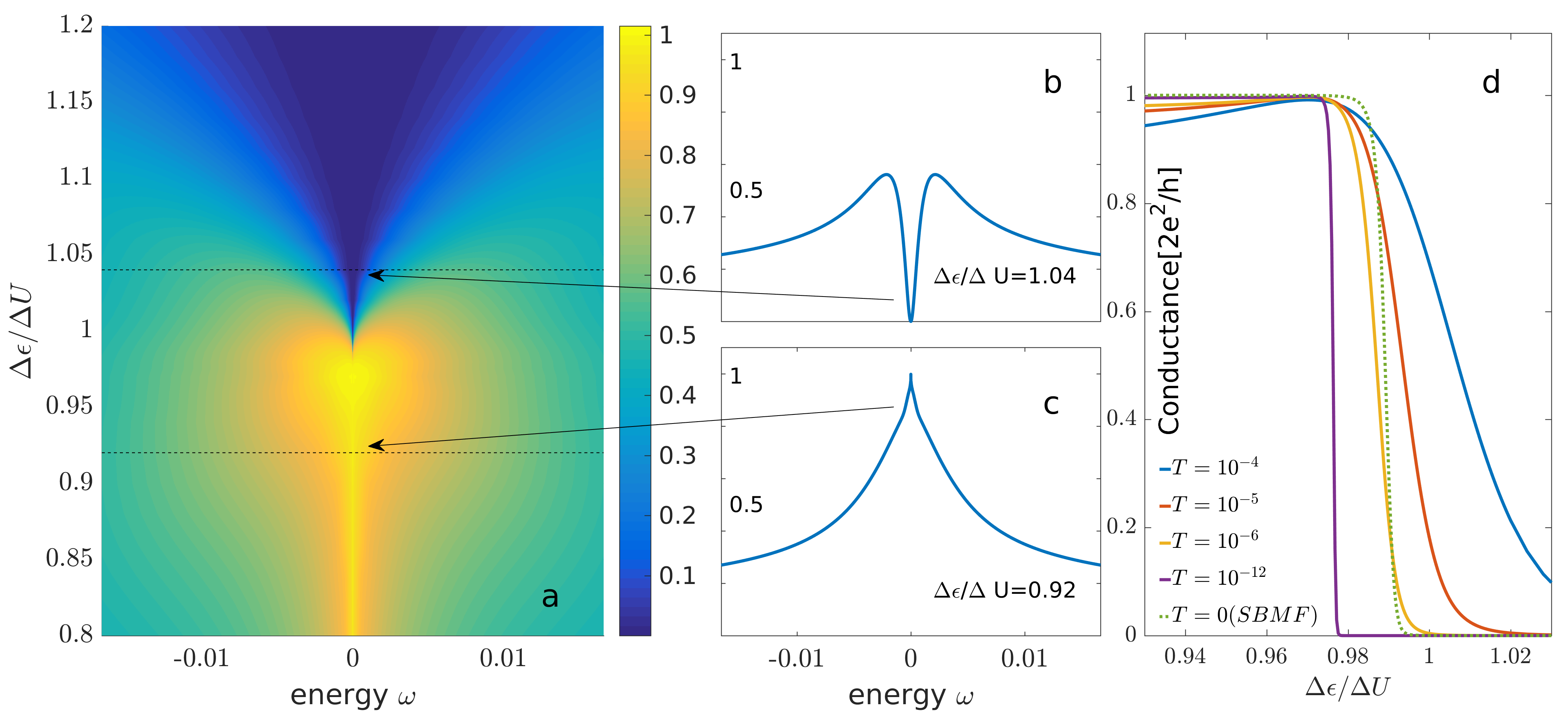}}
	\def\little{\mbox{\includegraphics[width=0.14\textwidth,height=0.06\textwidth]{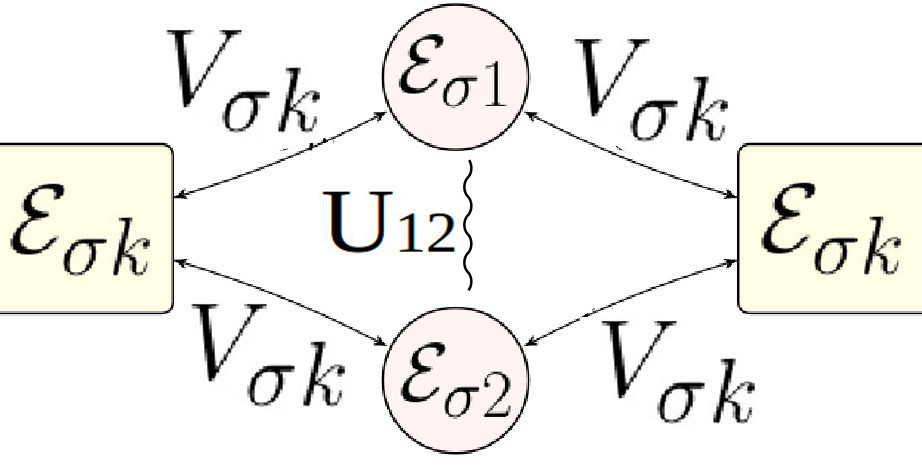}}}
	\def\stackalignment{l}
	\topinset{\little}{\big}{10pt}{35pt}
	
	\caption{(a) NRG results for the transmission spectral function, defined in the text, as a function of energy, and of $\de$, at temperature $T=3 \cdot 10^{-6}U$. Inset: The setup studied in this paper. (b,c) Transmission spectral function for two values of $\de$, denoted by the dotted line in (a),  as a function of energy. The Kondo peak for $\de/\dU<1$ seen in (b),  develops a dip in the vicinity $\de\gtrsim\dU$ (c), before the Kondo peak disappears altogether for $\de/\dU\gg1$.}
	\label{fig:DOS2d}
\end{figure}
 We first describe density-matrix numerical renormalization group (DM-NRG) results \footnote{We used the open-access Budapest Flexible DM-NRG
code, http://www.phy.bme.hu/dmnrg/; O. Legeza,
C. P. Moca, A. I. Toth, I. Weymann, G. Zarand,
arXiv:0809.3143 (2008) (unpublished)}. The  expectation values and the transmission spectral function (see below), required for the evaluation of the conductance through the double dot device \cite{Meir1992c}, were calculated, assuming, for simplicity, equal couplings to the left and right leads, $\Gamma=\pi \rho V^2$, and equal and constant density of states $\rho$ in the two leads, with a symmetric band of  bandwidth $D$ around the Fermi energy. In the following we set $U$ to be the unit of energy. The bandwidth value in the calculations is $D=3.33 U$, the intra- and inter- dot interaction difference is $\dU=U/6$ and the coupling to the leads is $\Gamma=U/15$.

\begin{figure}[h]
\includegraphics[width=0.32\textwidth,height=0.14\textwidth]{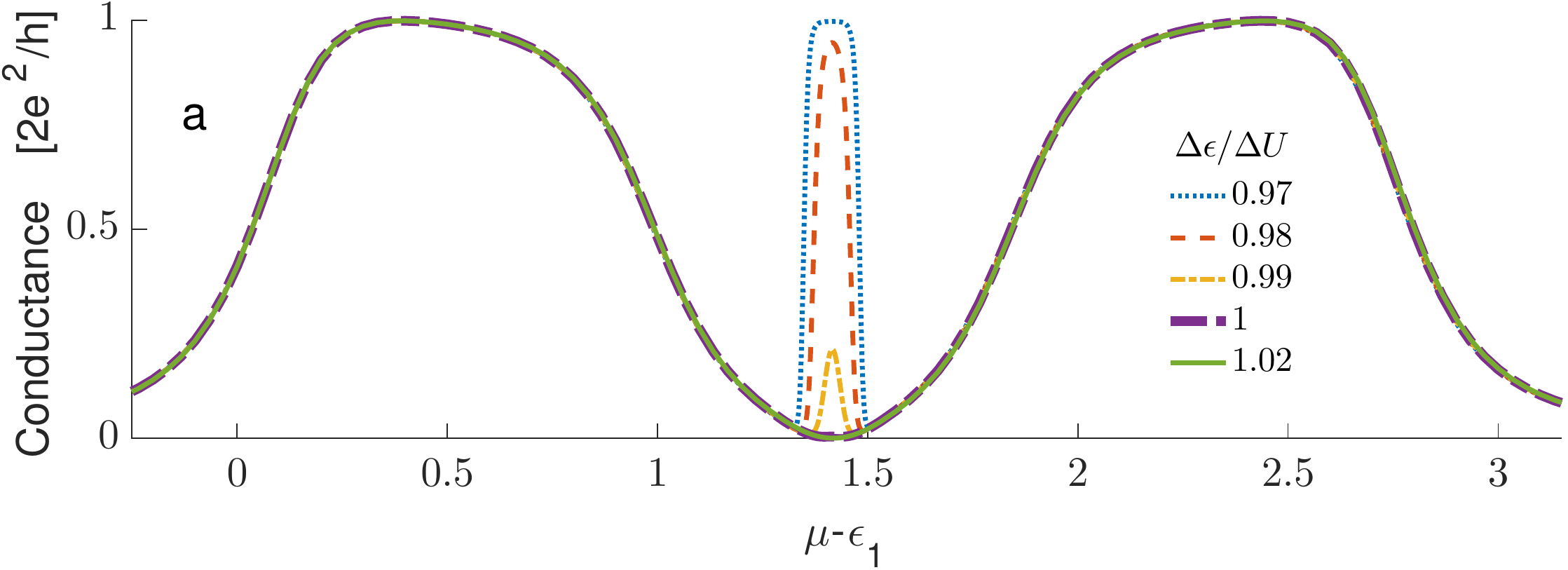}
\includegraphics[width=0.32\textwidth,height=0.14\textwidth]{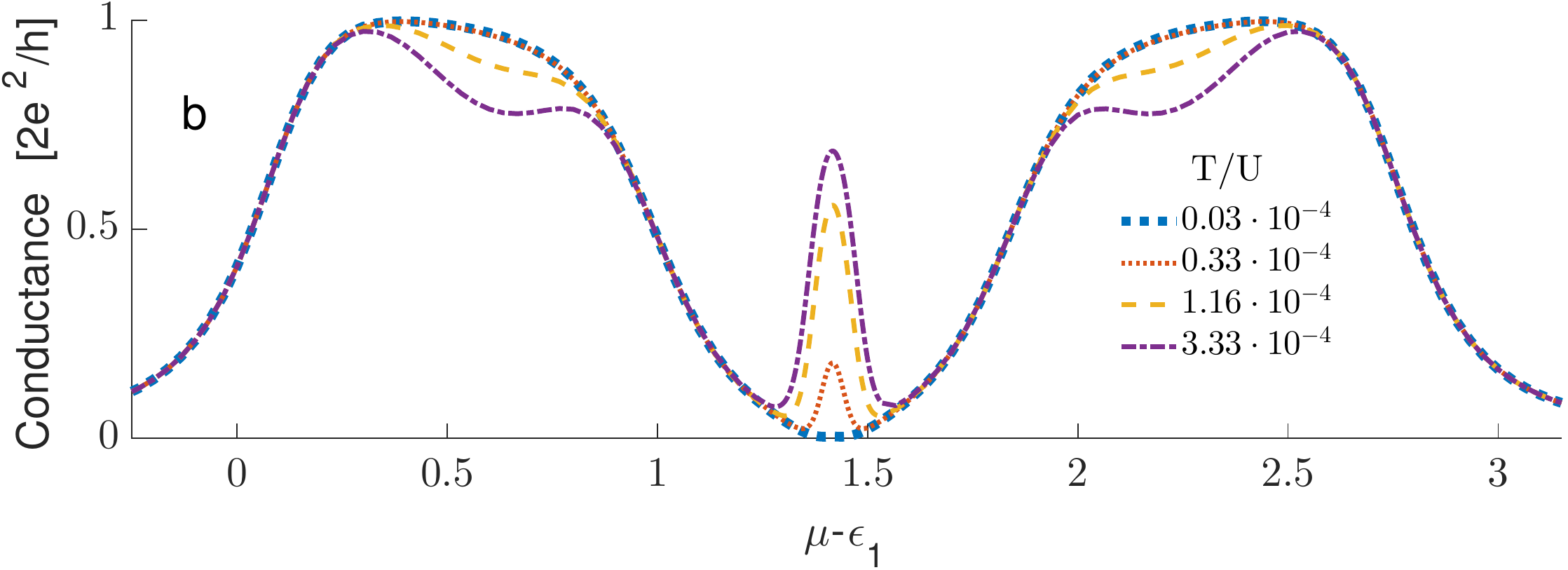}
\includegraphics[width=0.32\textwidth,height=0.14\textwidth]{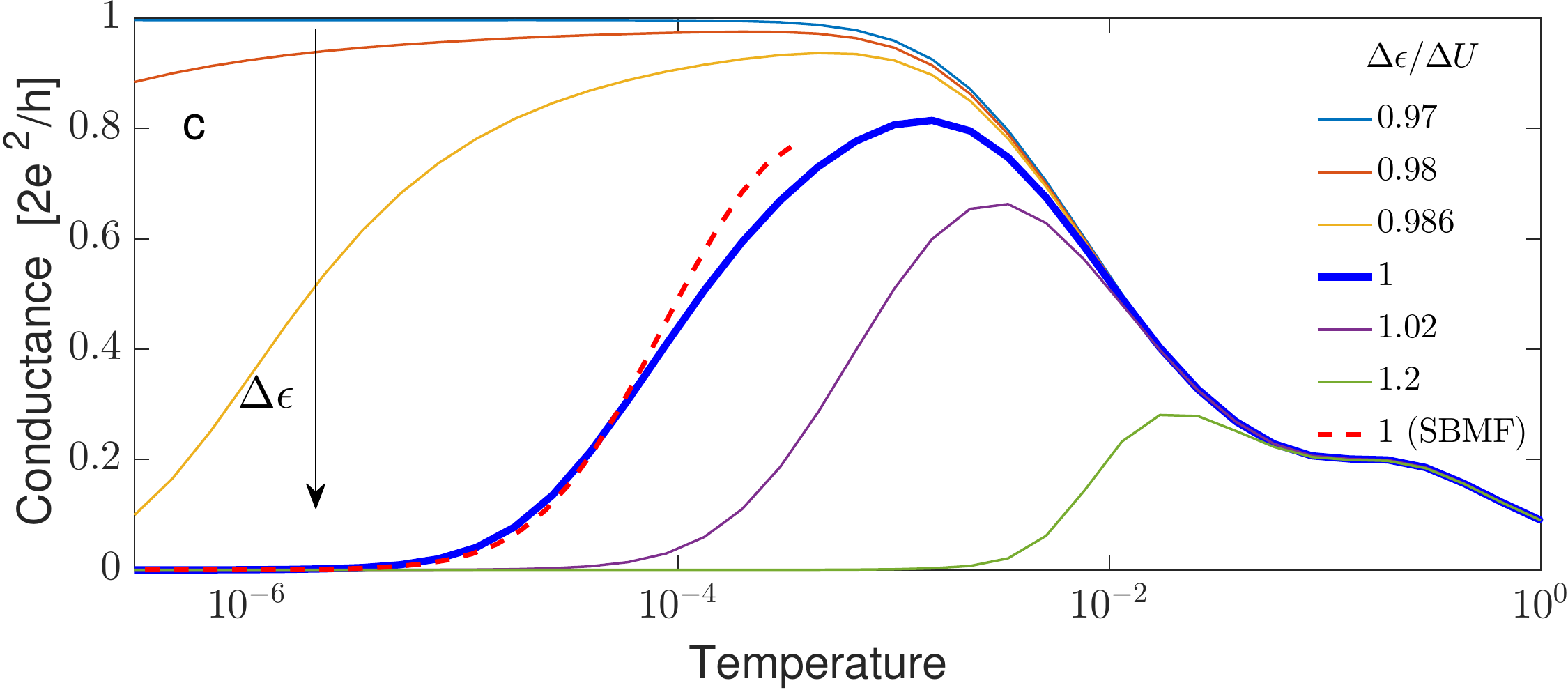}

 	\caption{NRG results for the conductance through the double-dot device as a function of chemical potential, for (a) various values of $\de$, at $T=3\cdot 10^{-6}$ (all energies are in units of $U$, the intra-dot interaction), and (b) for various values of temperature, with fixed $\de=\dU$. At zero temperature the conductance at the symmetry point $\mu=\mu_m$ is exactly zero, but a peak develops there at finite temperatures.  (c) The temperature dependence of the conductance at the particle-hole symmetry point, $\mu=\mu_{PH}$, for different values of $\de/\dU$. As one lowers the temperature the conductance starts to rise at $T\simeq T_{K1}$. For $\de/\dU\lesssim1$ the conductance rises and saturates at $G=2e^2/h$, in accordance with the underscreened Kondo effect. For $\de/\dU>1$ the conductance first rises around $T\simeq T_{K1}$, but then falls to zero for $T<T_{K2}$, demonstrating two-stage Kondo effect. While $T_{K1}$ seems independent of $\de/\dU$, $T_{K2}$ increases exponentially with $\de/\dU$. The broken line corresponds to the slave-boson mean field approximation, plotted in the low temperature range where it is more reliable.}
 	\label{fig:cond2ch}
 \end{figure}

Defining the  retarded Green functions $G^r_{ij,\sigma}(t-t')=-i\theta(t)\langle\{d_{i\sigma}(t),d_{j\sigma}(t')\}\rangle$ and the transmission spectral function $t_\sigma(\omega;\mu,T)=\frac{1}{\pi} Im[\Gamma\sum_{ij}\tilde{G}^r_{ij,\sigma}(\omega)]$, where $\tilde{G}^r_{ij,\sigma}(\omega)$ is the Fourier transform of $G^r_{ij,\sigma}(t)$, the current is given by \cite{Meir1992c}
\begin{equation}
J=\frac{e}{h}\sum_{\sigma}\int  t_\sigma(\omega;\mu,T) (f_L(\omega)-f_R(\omega))d\omega
\label{eq:current}
\end{equation}

Fig.~\ref{fig:DOS2d}a depicts a two-dimensional plot of transmission spectral function at the particle-hole symmetry chemical potential,  $\mu_{PH}\equiv\epsilon_1+(U+2U_{12}+\de)/2$, as a function of energy $\omega$ and energy difference $\de$ between the two dots,  for a fixed $\dU$ and low temperature $T=3\cdot 10^{-6}U$. The most striking feature of the data is the sharp change of behavior near $\de\simeq\dU$. For $\de\lesssim\dU$  (lower part of Fig.~\ref{fig:DOS2d}a) the transmission spectral function displays a sharp peak at the Fermi level (Fig.~\ref{fig:DOS2d}b), while for  $\de\gtrsim\dU$ there is a sharp dip at the Fermi level  inside a wider peak (Fig.~\ref{fig:DOS2d}c). The transition between these two regimes is very sharp: Fig.~\ref{fig:DOS2d}d displays the linear response conductance at that chemical potential (which is proportional to the  transmission spectral function at $\omega=0$), as a function of $\de/\dU$. We see that the conductance $G$ drops sharply from  $G=2e^2/h$ to nearly zero, indicating a quantum phase transition. In agreement with this interpretation, the critical regime becomes wider with increasing temperature \cite{Sondhi1997,Sachdev2011}. We will demonstrate below why this transition is indeed a QPT. The phase for $\de\ll\dU$ is relatively well understood: there are 4 degenerate states, as each dot is singly occupied with either spin. The states $|\uparrow,\uparrow>$, $1/\sqrt{2}\left(|\uparrow,\downarrow>+|\downarrow,\uparrow>\right)$ and $|\downarrow,\downarrow>$ form an $S=1$ entity (see also \cite{Zitko2006a}), while $1/\sqrt{2}\left(|\uparrow,\downarrow>-|\downarrow,\uparrow>\right)$ is an $S=0$ entity, that for equal coupling of the two dots to the leads, is, in fact, decoupled from the triplet (no tunneling through the leads). Thus this phase corresponds to the underscreened Kondo impurity\cite{Mehta2005,Hofstetter2002a}. 
 On the other hand, for $\de\gg\dU$ the first dot is doubly occupied, while the second one is empty, and thus there is no net magnetic moment on the double-dot system. Accordingly one would expect a small, finite contribution to the spectral function at $\omega=0$, and to the conductance, from the tails of the standard Coulomb blockade peaks. However, we find that the conductance there is exactly zero, within numerical accuracy. This is due to a  sharp dip at the transmission spectral function in the middle of a wider peak, that reaches all the way to zero (Fig.~\ref{fig:DOS2d}c). As we will show below, this is due to a two-stage screening mechanism.

  In order to further demonstrate the peculiar role played by the chemical potential $\mu_{PH}$, we plot in Fig.~\ref{fig:cond2ch}a the conductance as a function of chemical potential, for different values of $\de$, at the same temperature, $T=3\cdot 10^{-6}U$.
   Indeed, while for $\de<\dU$ the conductance exhibits a sharp Kondo peak in the middle of the double-occupation valley \footnote{A conductance peak at the PH symmetric point, similar in appearance to our case, also appears near the singlet-triplet transition, tuned by a magnetic field as studied in \cite{izumida2001}}, the conductance $G(\mu_{PH})$ vanishes there around $\de=\dU$, in accordance with the vanishing of the transmission spectral function at the chemical potential. Intriguingly,  as shown in Fig.~\ref{fig:cond2ch}b, as temperature is increased, $G(\mu_{PH})$ for that value of $\de$ increases, developing a sharp resonance, which is suppressed for yet higher temperatures. This might be expected from the form of the transmission spectral function: as the temperature increases beyond the width of the dip, the conductance should increase, and then decrease when the temperature becomes larger than the wider peak. The dependence of $G(\mu_{PH})$ on temperature, for various $\de$, is plotted in Fig.~\ref{fig:cond2ch}c. Such a non-monotonic dependence of the conductance on temperature, as seen for $\de/\dU\gtrsim 1$, is usually expected in the context of two-stage Kondo screening, with $T_{K1}$ and $T_{K2}$ the temperatures for the first and second stages. As the temperature $T$ is reduced below $T_{K1}$ the conductance starts to rise, but then falls towards zero for $T<T_{K2}$.
  As seen in Fig.~\ref{fig:cond2ch}, $T_{K2}$ becomes non-zero at $\de/\dU\simeq 1$, and increases sharply with increasing $\de/\dU$. Interestingly, for large enough values of $\de/\dU$, $T_{K2}$ becomes larger than $T_{K1}$ and the first stage does not fully form - the conductance decreases monotonically to zero with decreasing temperature due to local singlet formation.

\subsection*{Slave-Boson Mean-Field Theory}
\begin{figure}
 	\includegraphics[width=0.5\textwidth,height=0.17\textwidth]{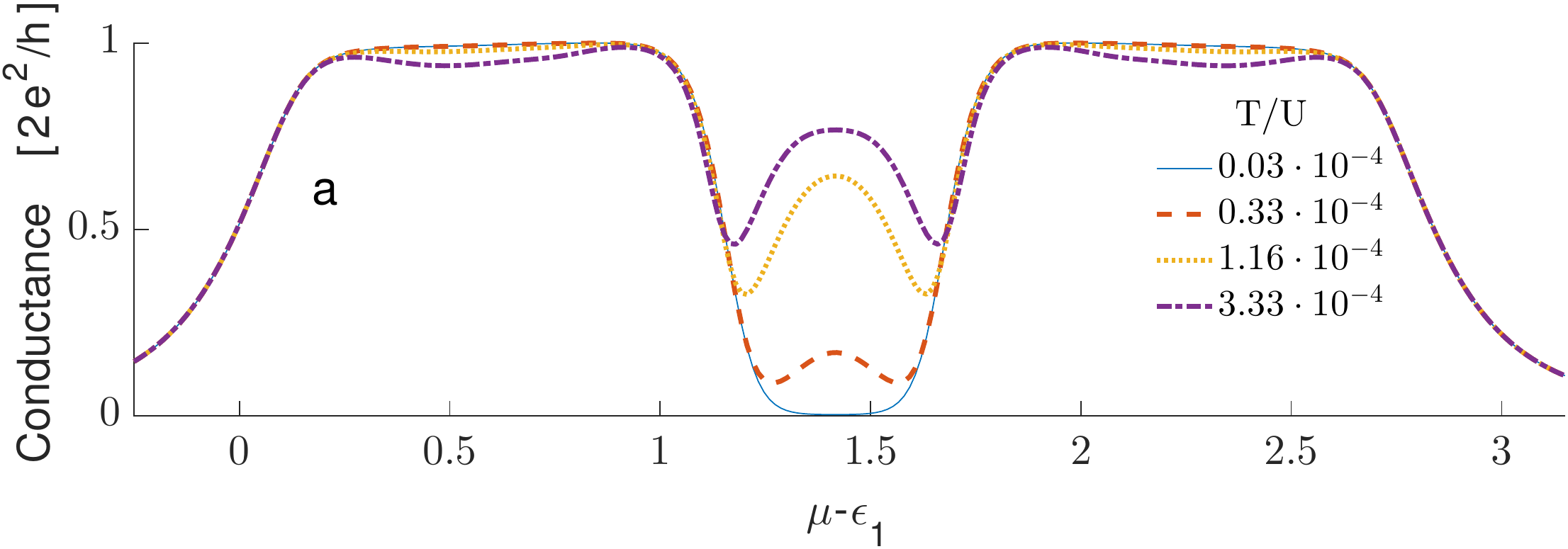}
 	\includegraphics[width=0.5\textwidth,height=0.15\textwidth]{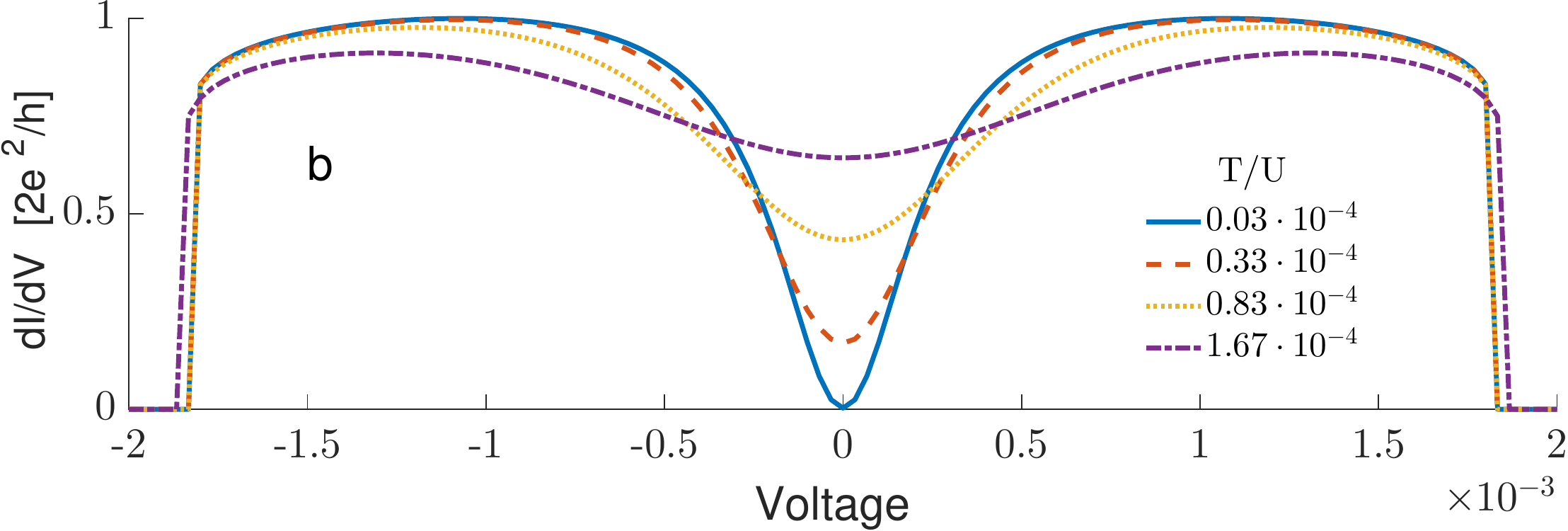}
 	\caption{Slave-boson mean-field results. (a) The linear-response conductance through the double-dot device as a function of chemical potential, for various values of temperature, with fixed $\de=\dU$. Other parameters are the same as in Fig.~\ref{fig:cond2ch}b, except $\Gamma$ which is smaller by a factor of 2 than the NRG calculation, $\Gamma_{SBMF}=U/30$. (b) The differential conductance $dI/dV$ at the particle-hole symmetric point as a function of voltage bias.}
 	\label{fig:SBMF}

 \end{figure}

In order to gain more insight into the physics behind the transition and in order to look beyond the linear response regime, we have employed the slave-boson mean-field approach in the Kotliar-Ruckenstein (KR) formulation \cite{Kotliar1986a}. In this method (see {\sl Methods}) one ends up with an effective non-interacting Hamiltonian, with renormalized parameters $\te_m$ and $\tilde{V}_m$, which, on average, obey the same constraints as the full interacting Hamiltonian.
The transmission spectral function for the effective non-interacting model
can then be easily expressed in terms of the renormalized parameters,
\begin{equation} t(\omega;\mu,T)\equiv \left|\frac{(\w-\te_1)\tG_2+(\w-\te_2)\tG_1
	}{(\w-\te_1+i \tG_1)(\w-\te_2+i \tG_2)+\tG_1 \tG_2}\right|^2
\end{equation}
where $\tG_m=\pi \rho \tilde{V}_m^2$ (note that even if $V_1=V_2$ in the original Hamiltonian, the renormalized parameters $\tilde{V}_m$ do not have to be the same when $\e_1\ne \e_2$). The temperature and chemical potential dependence of $ t(\omega;\mu,T)$ arise from the dependence of the renormalized parameters $\te_m$ and $\tG_m$ on these parameters. The resulting conductance, as a function of chemical potential, for the special point $\de=\dU$ is depicted in Fig.~\ref{fig:SBMF}a for $D=3.33U, \dU=U/6$ and $\Gamma=U/30$.  The results of the SBMFT approximation closely resemble those of the accurate NRG calculation (except for an overestimated width of the middle region), with the conductance going to zero at the symmetry point, only to increase with increasing temperature, giving rise again to a finite-temperature Kondo effect. The temperature dependence of the conductance at $\de=\dU$ in the SBMFT treatment is plotted along with the NRG results in Fig.~\ref{fig:cond2ch}c. The similarity between NRG and SBMFT results gives additional credence to this approximation, at least in this parameter regime \footnote{The above comparison was done for $\Gamma_{NRG}=2\Gamma_{SBMF}$, which is a known discrepancy between NRG and SBMFT \cite{Oguchi2010a}.}.

In addition to the features in the linear-response conductance, the predicted dip in the spectral function can be probed by measuring the voltage-dependent differential conductance $G(V)$ through the double-dot system for  $\de=\dU$.  Fig~ \ref{fig:SBMF}b depicts  $G(V)$ for several temperatures, using the SBMFT approximation. As one might expect, $G(V)$ exhibits a dip at zero bias, corresponding to the shift of the peaks in the spectral function from the Fermi energy. At high enough voltage the Kondo effect is suppressed, though in the SBMFT approach, this appears as an unphysical abrupt transition \cite{Aguado2000b}.

 Within SBMFT, the emergence of a Kondo peak in the spectral function, and the resulting zero-bias anomaly in the conductance, are due to the renormalization of the effective energies $\te_i$ toward the chemical potential. For a single dot, in the Coulomb-blockade valley corresponding to total unit occupation, each spin state is half-occupied, on average. Thus, in order to obtain the same occupation by an effective non-interacting model, the energy of each spin state lies exactly at the Fermi energy, leading to a resonance at that energy which is interpreted as the  Abrikosov-Suhl resonance associated with the Kondo effect (for a review, see \cite{Hewson}). On the other hand, in the single-dot Coulomb-blockade valleys which correspond to either zero or double occupation, the renormalized energy levels are shifted to well above or below the Fermi level, leading to suppression of the spectral function at the Fermi energy. Thus, in order to understand the features in the spectral function in the double-dot system, one needs to determine the corresponding energy shifts.  For $\de<\dU$ the two dots are singly occupied, and thus their energies are degenerate, and, as in the single-dot case, are shifted to the chemical potential, leading to a peak at the spectral function at that energy. On the other hand, for $\de\geq\dU$, the occupation of dot $1$ is larger than that of dot $2$, so the energies $\te_1$ and $\te_2$  straddle the Fermi energy symmetrically (since the occupation has to add up to $n=2$). Moreover, the two symmetric Abrikosov-Suhl resonances on the two sides of the Fermi energy give rise to the exact same transmission amplitude, which interfere destructively due to a phase difference of $\pi$ between sub-resonance and sup-resonance transmission through the individual dots. This interference is the origin of the central dip and the finite-temperature effect. Interestingly, exactly at the transition point, where $\de=\dU$, the occupations of the two dots in the
 SBMFT treatment are $n_1=6/5$ and $n_2=4/5$, so the renormalized energies assume specific values in the mixed-valence regime, but the conductance is still exactly zero during the above-mentioned interference effect.

The SBMFT formulation offers an alternative point of view of the physics of the transition. Concentrating on the PH symmetry point, $\mu=\mu_{PH}$, the constraints of the SBMFT equations require $\te_1=-\te_2$ and $\tilde{V}_1=\tilde{V}_2$. Transforming into the even-odd basis, $d_{\bfrac{e}{o}\sigma}=\frac{1}{\sqrt{2}}\left(d_{1\sigma}\pm d_{2\sigma}\right)$, the effective SBMFT Hamiltonian takes  the following form:
 \begin{equation}
 H_{MF}=\sum_\sigma(\te_1 d^\dagger_{e\sigma}d_{o\sigma}+h.c)+
 \sum_{\sigma k\in L,R} \epsilon_{k}c^\dagger_{k\sigma} c_{k\sigma}+\sum_{\sigma,k\in L,R}(\sqrt{2}\tilde{V}c^\dagger_{k\sigma}d_{e\sigma}+h.c.)
 \label{eq:EO_SB}
 \end{equation}

So in the even-odd language, the leads are only coupled to the even state, with possible hopping between the even and  the odd state, which is proportional to $\te_1$. As was shown above, for $\de<\dU$ the SBMFT equations lead to $\te_1=0$, resulting in a single state  coupled to the leads, exactly on resonance, which is the standard SBMFT Kondo solution. On the other hand, for  $\de\geq\dU$ $\te_1$ becomes finite and grows, thus allowing for tunneling between the even and odd states, resulting in splitting of the energies symmetrically around the Fermi energy. This is identical to the results of the SBMFT calculation for side-coupled quantum dot (depicted in the inset to Fig.~\ref{fig:SS}b) \cite{Cornaglia2005}, where $\de-\dU$ plays the role of the effective magnetic interaction between the dots. Thus in this language, the QPT discussed above is expressed in terms of the standard Kondo transition from a FM to AFM interaction \cite{Anderson1970c}.


\subsection*{The Hamiltonian in the language of even-odd states}

\begin{figure}
		\includegraphics[width=0.5\textwidth,height=0.17\textwidth]{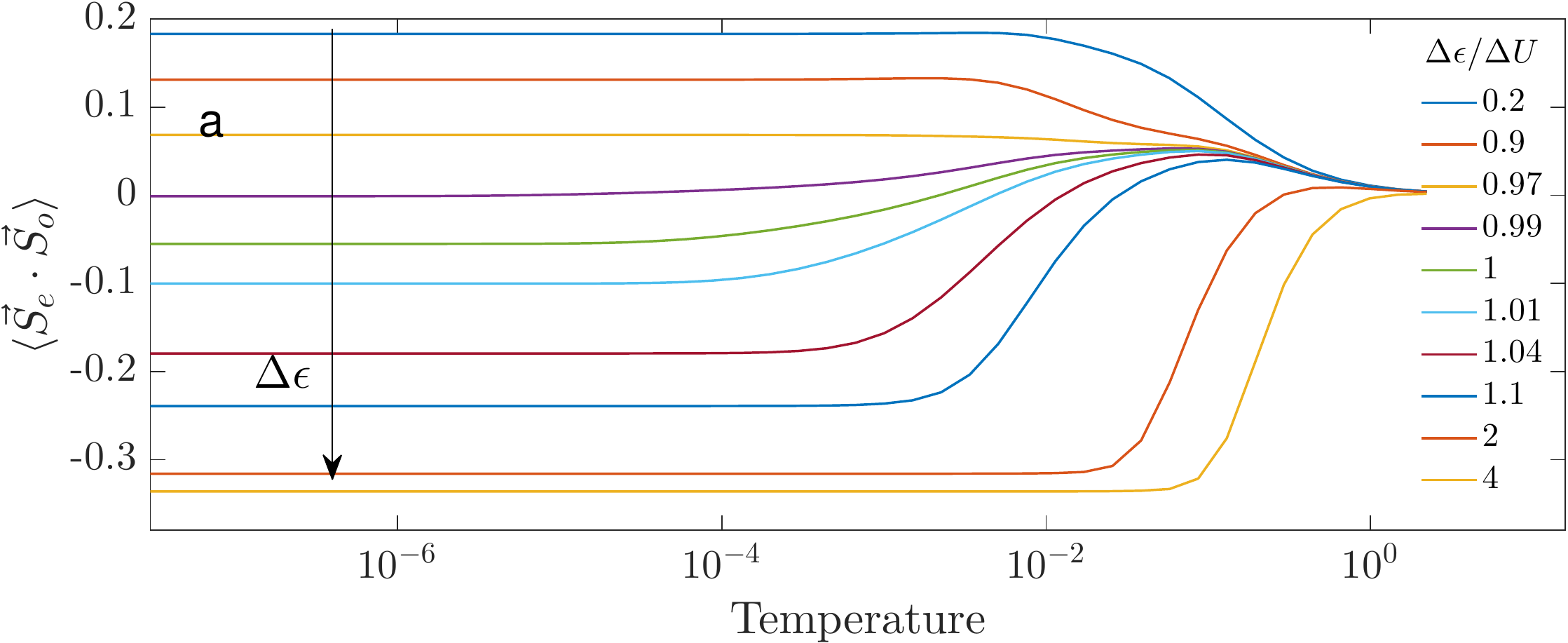}
	\def\big{	\includegraphics[width=0.35\textwidth,height=0.17\textwidth]{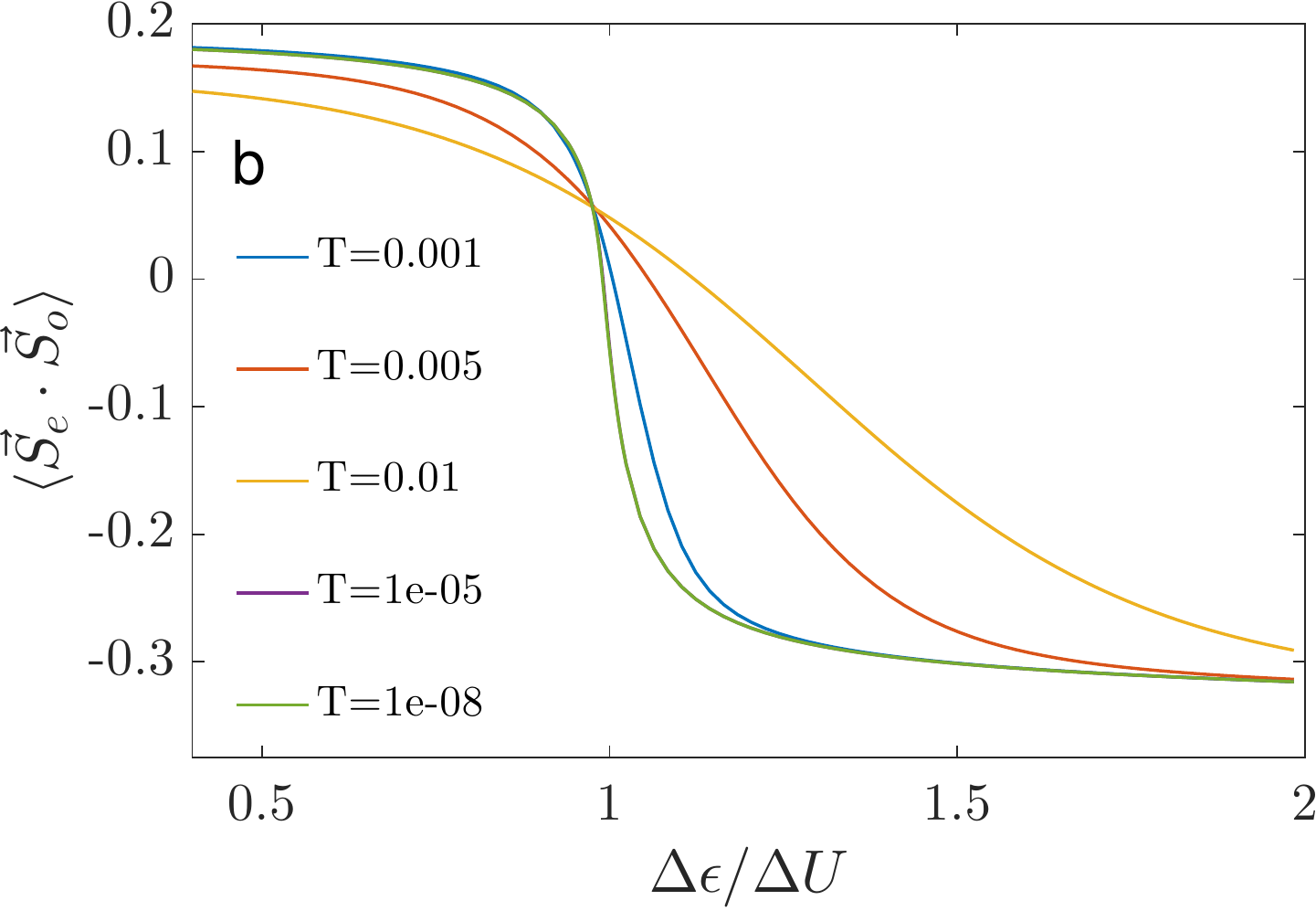}}
		\def\little{\mbox{\frame{\includegraphics[width=0.20\textwidth,height=0.10\textwidth]{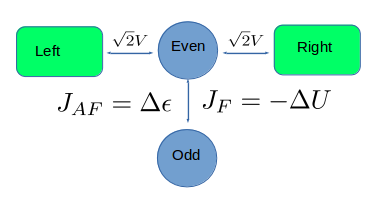}}}}
		\def\stackalignment{l}
		\topinset{\little}{\big}{-5pt}{120pt}
			\caption{NRG results for the spin-spin correlation function in the even-odd basis (a) as a function of temperature for different values of $\de$, (b) as a function of $\de$ for various temperatures, demonstrating a transition from ferromagnetic to anti-ferromagnetic correlations with changing $\de/\dU$. Inset: the effective system in the even-odd language. Only the even "dot" is coupled to the leads, and $\de$ and $\dU$ generate ferromagnetic and antiferromagnetic couplings between these dots, respectively.}
			\label{fig:SS}
	\end{figure}
Motivated by the insight provided by the SBMFT results, we rewrite the Hamiltonian (Eq.~\ref{eq:H2d}) in the even-odd basis,
\begin{equation}
H_{2QD}=\sum_{p \sigma} \e_{av} \hat{n}_{p \sigma}+
\sum_{\sigma} t( d^\dagger_{e \sigma}d_{o \sigma}+ h.c)+ \sum_{p} \tilde{U} \hat{n}_{p\uparrow}\hat{n}_{p\downarrow} + \tilde{U}_{12} \hat{n}_e \hat{n}_o+J_F\vec{S}_e\cdot\vec{S}_o +A(P^+_\uparrow P^+_\downarrow+P^-_\uparrow P^-_\downarrow)
\label{eq:H2deo}
\end{equation}
where the parameters of the new Hamiltonian are $\e_{av}=\frac{1}{2}(\e_1+\e_2), t=\frac{1}{2}\de$, $\tilde{U}=U-\dU/2, \tilde{U}_{12}=U-3\dU/4, J_F=-\dU$, and $A=\dU/2$. The parity ladder operators are defined $P^{+(-)}_\sigma=c_{e(o),\sigma}^\dagger c_{o(e),\sigma}$ and will contribute very little to the dynamics in the given range.  The full Hamiltonian, including tunneling and leads becomes
\begin{equation}
H = H_{2QD}\ + \sum_{\sigma k\in L,R} \epsilon_{k}c^\dagger_{k\sigma} c_{k\sigma}+ \sum_{m \sigma k\in L,R} \left(\sqrt{2}V d^\dagger_{e \sigma} c_{\sigma k } + h.c\right) ,
\label{eq:H2deoFull}
\end{equation}

which shows, as before, that the odd level is entirely decoupled from the leads. While most of the specific values are of little qualitative importance, there are two new terms which draw the most interest. The hopping amplitude $t$ gives rise to an effective AFM interaction $J_{AF}= 2t=\de$\footnote{unlike the usual side coupled effective interaction$J_{AF}\sim \frac{4t^2}{\dU}$, here the extra terms of $J_F=-\dU$ and $A=\dU/2$ change the spectrum so that $J_{AF}=2t=\de$.} in the $n_e=n_o=1$ subspace, while the interaction $J_F$ is a FM one. As a result, the effective Kondo model would represent a conduction spin coupled AFM to a  localized spin which is further coupled to another spin via $J_{tot}=J_{AF}+J_F=\de-\dU$  \footnote{There is an additional n=2 level which is degenerate with the spin-triplet state. This level does not incur spin-flip processes through the leads and does not contribute to the Kondo screening.} The competition between $J_F$ and $J_{AF}$ governs the quantum phase transition between the two phases mentioned. This analysis is supported by Fig.~\ref{fig:SS}, which shows the spin-spin correlation between the even and odd states,  an indicator for the sign of the magnetic interaction and its magnitude. The transition from FM to AFM occurs a little below $\de/\dU\simeq1$ due to level renormalization through the continuum, a temperature dependent effect.  For FM $J_{tot}<0 (\de<\dU)$ the two localized spins form an $S=1$ triplet, leading to an under-screened Kondo impurity \cite{Vojta2002} (with an additional, uncoupled level  with $S=0$, which gives rise  to  $\langle \vec{S}_e\cdot\vec{S}_o\rangle=0.25\cdot \frac{3}{4}$).
As $\de$ increases further $\langle \vec{S}_e\cdot\vec{S}_o\rangle$ become negative, eventually saturating at $0.75\cdot \frac{1}{2}$, as the ground state for $\de\gg\dU$ is an equal superposition of the two singlets, $\frac{1}{\sqrt{2}}\left(|\uparrow\downarrow,0\rangle-|0,\uparrow\downarrow\rangle\right)$ and $\frac{1}{\sqrt{2}}\left(|\uparrow,\downarrow\rangle-|\downarrow,\uparrow\rangle\right)$. The transition from positive to negative spin-spin correlation function  demonstrates that indeed $\de-\dU$ plays the role of the magnetic interaction which drives the system between the two phases.

\section*{Discussion}
We have demonstrated in this paper that for a realistic double quantum-dot device, one can tune the system through a quantum phase transition, leading to a sharp change in the conductance and in the shape of the zero-bias anomaly.
The transition between the reported phases and their respective conductance signatures have been studied here as two facets of the same effect. On one hand it is described by the destructive vs constructive interference between the two dot branches.
On the other side it is described by the transition between ferro- or anti-ferromagnetic interaction between two impurities in a side coupled setup, which are relatively well understood. The predicted features in the conductance, either as a function of gate voltage (Fig.~\ref{fig:cond2ch}a and b) or as a function of temperature (Fig.~\ref{fig:cond2ch}c), can be easily checked in the double-dot setup, depicted in the inset of Fig.~\ref{fig:DOS2d}, where each dot is controlled by a different gate voltage. The intra- and inter-dot interactions, $U$ and $U_{12}$, and consequently $\dU=U-U_{12}$, are usually determined by the geometry of the system and cannot be easily modified. However the energy difference between the two dots, $\de=|\e_1-\e_2|$, can be readily tuned. Thus, for a given setup, one can change this relative voltage until the Kondo peak at the valley midpoint is suppressed and the conductance vanishes. As discussed above, this should happen around the point where $\de$ reaches the value $\dU$. While in this paper we have concentrated on the single-channel case, where both dots are connected to the same channel in the leads, one can easily extend the calculation to  the general case of mixed channels, where the lead states the two dots couple to are not identical. Even in this case one expects that the transition studies here   will also give rise to an observable effect as long as these lead wave-functions are not orthogonal. Thus, we hope that the results presented here will stimulate experiments in this direction.

\section*{Methods}
 Within the KR slave-boson framework\cite{Kotliar1986a}, the Hamiltonian is replaced by an effective non-ineracting Hamiltonian where slave bosons are added as an accounting tool for the different many-body states of the system.  In order to accommodate the larger Hilbert space of the two-dot Hamiltonian, an enlarged boson space is introduced \cite{Oguchi2010a}, where each boson accounts for one of the impurity dimer states: $e$ for empty dimer, $p_{\sigma m}$ for single occupation of an electron with spin $\sigma$ on dot $m$, $x_m$ for two electrons on dot $m$, $y_{s \ell}$ for two electrons on different dots with total spin $s$ and z-component of total spin $\ell$, $h_{\sigma m}$ for triple occupation, with a missing electron on dot $m$ with spin $\sigma$, and $b$ for four electrons in the dimer. Since the system can be in only one of these states (or a linear combination with total weight equal to unity), the bosons are constrained to have unit total occupation $I=e^\dagger e+\sum_{\sigma m}p^\dagger_{\sigma m}p_{\sigma m}+\sum_i x_i^\dagger x_i+\sum_{s l}y_{sl}^\dagger y_{sl}+\sum_{\sigma m}h^\dagger_{\sigma m}h_{\sigma m}+b^\dagger b$. Given the boson operators, the electron number operators are constrained to be $\hat{n}_{\sigma m}=\hat{Q}_{\sigma m}$, where  $\hat{Q}_{\sigma m}=e^\dagger e+p^\dagger_{\sigma m}p_{\sigma m}+\sum_m x_m^\dagger x_m+y_{1 2\sigma}^\dagger y_{1 2\sigma}+\frac{1}{2}\sum_{s}y_{s 0}^\dagger y_{s 0}+h^\dagger_{-\sigma m}h_{-\sigma m}+\sum_{\sigma}h^\dagger_{\sigma -m}h_{\sigma -m}+b^\dagger b$. Including these constraints, the Hamiltonian (\ref{eq:H}) can be exactly mapped into the slave-boson Hamiltonian, $H_{SB}=H_{2QD}+H_{lead}+H_{cons}$, where:
 \begin{align}
 H_{2QD}&=\sum_{\sigma,m}\epsilon_{m}\hat{n}_{m\sigma}+U\sum_{m}x_m^\dagger x_m+U_{12}\sum_{s l}y_{s l}^\dagger y_{s l}
 +(U+2U_{12})\sum_{\sigma m}h^\dagger_{\sigma m}h_{\sigma m}+(2U+4U_{12})b^\dagger b\\
 H_{cons}&=\sum_{\sigma m}\lambda_{\sigma,m}(\hat{n}_{m\sigma}-\hat{Q}_{\sigma m})+\lambda(I-1)\nonumber \\
 H_{lead}&=\sum_{\sigma k\in L,R} \epsilon_{k}c^\dagger_{k\sigma} c_{k\sigma}+\sum_{\sigma,m,k}z_{m\sigma}V_{m}c^\dagger_{k\sigma}d_{\sigma m}+h.c.\nonumber
 \label{eq:HSB}
 \end{align}
 where $z_{m\sigma}=(1-\hat{Q}_{\sigma,m})^{-1/2}(e^\dagger p_{\sigma m}+p^\dagger_{-\sigma m}x_m+p^\dagger_{\sigma -m}y_{1 2\sigma}+p^\dagger_{-\sigma -m}(y_{0 0}+y_{1 0})/2+x^\dagger_{-m}h_{m -\sigma}+(y^\dagger_{0 0}+y^\dagger_{1 0})h_{-\sigma -m}/2+y^\dagger_{1 2\sigma}h_{\sigma -m}+h_{\sigma m}^\dagger b)(\hat{Q}_{\sigma,m})^{-1/2}$.
 In the mean-field approximation, all the boson operators are replaced by their expectation values, and one solves for these values using the self-consistent equations of motion and constraints (see Ref.~\cite{Oguchi2010a} for further details). As a result of the mean-field approximation, one ends up with an effective non-interacting model, where the parameters of the two dots are renormalized ($\e_{m}\rightarrow\te_m\equiv\e_{m}+\lambda_{\sigma m}$ and $V_{m}\rightarrow\tilde{V}_{m}\equiv z_{m\sigma}V$), where we assume spin degeneracy, so the bosonic expectation values are independent of $\sigma$.

\bibliography{SecondAnomaly}

\end{document}